# TIC como apoyo del soporte social al enfermo crónico y su cuidador: Aproximación al estado del Arte

**Andres F. Ardila, Pedro L. Cifuentes, Benjamin A. Huerfano y Marcia C. Pulido**

Resumen

El enfoque actual se lleva a cabo con el fin de tener un panorama sobre el nivel de inclusión y la participación de las TIC en el soporte social y el apoyo a las poblaciones vulnerables que sufren de enfermedades de carácter crónico. Se hizo la inclusión a través de una revisión bibliográfica, siendo esta la base de la recopilación de datos e información pertinente. El estudio argumentativo que se llevó a cabo, identificó de forma clara y concisa las ventajas y desventajas del uso de las TIC en el soporte social desde un punto de vista psicoeducativo e ingenieril. La regiones se caracterizaron por la mayor concentración de uso de las TIC en la literatura de apoyo social, basado en el contenido previamente estudiado y analizando los resultados de este uso.

*Palabras Clave*— Soporte social, TIC, Enfermo Crónico, Cuidador.

## I. INTRODUCCIÓN

La cotidianidad de la vida humana, resalta como un aspecto de vital importancia la *salud*, siendo esta inestable y dependiente de muchos aspectos, tales como emocionales, físicos, cognitivos, sociales, entre otros. Además se puede definir el aspecto de la salud como presencia o ausencia de enfermedad, donde esta primera crea una gran vulnerabilidad de la persona que padezca dicha enfermedad, afectándolo en el modo en que se desenvuelve dentro de la sociedad. Actualmente el aumento demográfico de enfermos crónicos, ha evidenciado que es una problemática de salud a nivel mundial con casi un 72% del total de la carga de morbilidad en personas mayores de 30 años [1] y esto se hace evidente debido al vuelco poblacional que se ha venido presentando en las últimas décadas, que muestra un aumento en el número de adultos mayores haciendo más prevalente la aparición de enfermedades crónicas a medida que la población envejece [2]. Teniendo en cuenta lo descrito, se observa claramente la necesidad de aplicar redes de soporte social para el cuidado de enfermos crónicos.

Hoy en día, las Tecnologías de Información y las Comunicaciones (TIC), se han convertido en una herramienta de gran importancia en todos los campos de las distintas áreas del conocimiento. El surgimiento de las nuevas tecnologías ha cambiado en las personas la forma de trabajar, divertirse, relacionarse, hasta de aprender y cuidar la salud [3]. De esta necesidad surge, la idea de hacer uso de dicha evolución tecnológica para poder generar bienestar dentro de una sociedad y aplicarlas dentro de las redes de apoyo, en este caso las encaminadas al soporte social de enfermos crónicos y sus cuidadores, puesto que no es suficiente el apoyo brindado por esta redes, sin ayuda de la tecnología, haciendo plausible el uso de nuevos medios de comunicación dentro de las redes de soporte social, para ampliar su cobertura; por otra parte se ha identificado dificultades en el acceso de los servicios especializados de salud y al equipo profesional sensibilizado con dicha problemática, ocasionando así un desaprovechamiento de las tendencias y los recursos actuales.

## II. SOPORTE SOCIAL

Aunque el concepto de soporte social nació algunas décadas atrás aproximadamente en los años 60, fue hasta los 70 cuando comenzó con mayor fuerza [4]. La tendencia inicial de los estudios empíricos fue considerar el soporte social, como un constructo unidimensional y analizarlo a partir del ámbito en que se desenvolviera dicho soporte [5]; Existe muchos puntos de vista acerca del soporte social y como tal se ha definido en términos de actividades, comportamientos, relaciones e interacciones o también en términos de la calidad de las relaciones sociales. A pesar de la diversidad de puntos de vista existe un punto de consenso según el cual se define al soporte social, como una estructura compleja y multidimensional que tiene en cuenta de manera conjunta la calidad de las relaciones sociales, para mejorar de manera directa un bien o estilo de vida.

Históricamente hay dos disciplinas que se han centrado en la contribución, popularización y desarrollo del soporte social, las cuales son la epidemiología y la psicología; la primera se ha centrado en el impacto del tejido social sobre la prevalencia y mejoría de algunas enfermedades y las tasas de mortalidad por estas mismas y la última ha aportado evidencia del impacto de tipo benéfico del soporte social sobre los individuos [6], [7].

Andres F. Ardila: afardila@mail.unicundi.edu.co, estudiante de Ingeniería Electrónica, Universidad de Cundinamarca - Colombia.

Pedro L. Cifuentes: pedroelectric2005@gmail.com, Docente investigador de ingeniería Electrónica, Universidad de Cundinamarca - Colombia

Benjamin A. Huerfano: bhuerfano@mail.unicundi.edu.co, estudiante de Ingeniería Electrónica, Universidad de Cundinamarca - Colombia.

Marcia C. Pulido: catapul14@gmail.com, Docente investigador de ingeniería Electrónica, Universidad de Cundinamarca - Colombia

El soporte social puede fácilmente ser categorizado en dos distintas subdivisiones o ramas, la formal y la informal, de las cuales cabe mencionar que tienen valores agregados y/o son complementarias. El principal pilar del soporte social se encuentra dentro de la rama del informalismo, siendo el núcleo familiar la base de la administración de cuidados del paciente crónico; sin su participación activa difícilmente se logran cuidados adecuados. La familia tiene un papel apaciguador de todas las tensiones que se van creando a lo largo del proceso de la enfermedad crónica, sin importar que la familia se adapte adecuadamente a la situación. La enfermedad crónica conlleva a una brecha y una crisis de la vida cotidiana, implicando un reajuste de los miembros de dicho núcleo. Según Alvarado (2010), "del 70% al 80% de cuidado brindado a personas con situaciones de enfermedad crónica es promovido por la familia" [8]; lo cual reafirma el hecho de que la familia es la base de un buen soporte social.

Las redes formales, en cambio, se han establecido con el propósito específico de dar apoyo a las personas adultas mayores afectadas por una enfermedad. En éstas, sus miembros cumplen roles concretos y algunas veces requieren contar con la preparación adecuada. Las redes formales pueden ser parte de centros comunitarios, de salud, o de programas gubernamentales [9]. A comparación del soporte social informal, su antónimo brinda una perspectiva más directa y concisa de las tareas que desarrolla el soporte social, como lo hacen las instituciones hospitalarias y programas de salud, que buscan brindar apoyo profesional en las zonas con mayor vulnerabilidad [10]. En estos escenarios se deben implementar acciones o programas encaminados a la promoción de la salud, no sólo en beneficio de los enfermos sino de las personas que los cuidan (cuidadores), ya que ellos necesitan capacitación y estímulos para el manejo del enfermo o apoyo para su propio autocuidado.

Numerosos datos apoyan la idea de que las personas que cuentan con buenas redes sociales se adaptan fácilmente a su situación, lo cual protege contra el estrés generado por la enfermedad y capacita al paciente para reevaluar la situación y adaptarse mejor a ella, ayudándole a desarrollar respuestas de afrontamiento a dicha enfermedad. [9], [11]-[13].

El soporte social se compone en pocas palabras de un apoyo directo en el desenvolvimiento de actividades o un objetivo en específico por parte de un grupo social, organización o programa y de esta manera generar un aporte económico, logístico, organizacional, psicológico, emocional, o cualquier otro.

### A. Enfermo crónico

En la literatura se encuentran múltiples definiciones acerca del enfermo crónico, ya que lo pueden definir desde diferentes perspectivas debido a que es disímil ver un enfermo crónico con los ojos de un profesional de la salud, a verlos con los ojos de una persona con vínculo o parentesco. En este caso se trata de definir al enfermo crónico como la persona que necesita cuidado. El enfermo crónico es la persona que presenta una enfermedad que no se cura, pero se puede manejar mediante la modificación de los modos de vida, tratamientos oportunos y cuidados permanentes según su situación de salud. La enfermedad crónica perdura a través del tiempo, y esto genera un deterioro progresivo, que en algunos casos lleva a que la persona presente un nivel de dependencia y requiera de un cuidador.

### B. Cuidador

Existen varios tipos de cuidadores que pueden resumirse en: formales que son los profesionales de la salud y los informales que en su mayoría son familiares, que son aquellas personas adultas "con vínculo de parentesco o cercanía afectiva que asume las responsabilidad del cuidado de la persona con enfermedad crónica, lo apoya en las actividades de la vida diaria, y participa con él en la toma de decisiones sobre su cuidado"[14] . Desde hace algún tiempo, se ha encontrado que éstos sufren el llamado "Síndrome de cuidador", el cual se define como "una respuesta inadecuada a un estrés emocional crónico cuyos rasgos principales son un agotamiento físico y/o psicológico, una actitud fría y despersonalizada en la relación con los demás y un sentimiento de inadecuación a las tareas que ha de realizar" [15], lo cual afecta gravemente su calidad de vida.

### C. Relación

Dentro de este marco referente a enfermedades crónicas se ven implicados diferentes contextos tales como sociales, físicos, emocionales, tanto del cuidador como del enfermo crónico y de sumisión de este último en cuanto a su cuidado. A través del estudio de estas implicaciones, se han generado múltiples estrategias como son de tipo educativo, psicoeducativas, de soporte, psicoterapéuticas, de apoyo con el cuidado, de entrenamiento para el receptor del cuidado y la persona que es atendida, que fomenta la autoayuda de los cuidadores familiares, como un camino para remediar la situación descrita [16] y contrarrestar los efectos negativos de la enfermedad. Estas estrategias que se han desarrollado traen consigo unas deficiencias, que tiene que ver con la accesibilidad de los cuidadores y enfermos crónicos a estos nuevos aprendizajes y a la información, ya que el enfermo crónico depende de su tutor en todo momento, sumándole que no pueden acceder al programa desde su casa, esto dificulta la introducción a un programa de formación.

Cabe resaltar que el cuidador informal "debe adquirir habilidades y conocimientos para poder prestar la atención requerida por el paciente"[17] y se hace indispensable para mejorar la calidad de vida. Debe contar con la habilidad de cuidado que "se constituye en el arte, la pericia, la maestría y la experiencia de dar cuidado de manera tal que satisfaga los requerimientos de la persona a quien se cuida"[18]. Algunos deberes propuestos para el cuidador están relacionados con la necesidad de informarse sobre la enfermedad y adquirir un mayor aprendizaje a través de la experiencia, en aspectos como la organización de los recursos, las acciones de cuidado, el manejo del tiempo y la apropiación de su labor. Lo anterior implica que los cuidadores "deben ser capaces de superar dificultades y presiones, enfrentar el desconocimiento y el miedo al cuidado, aprender a afrontar los momentos difíciles que se generan, producto de la conducta desordenada del enfermo, y de las situaciones propias del cuidado como la monotonía, el aislamiento y la falta de apoyo." [19]

### III. REDES DE SOPORTE SOCIAL

Los países han estado reformando su sistema de salud debido al cambio demográfico que se ha venido presentando. El grupo de profesionales no crece lo suficientemente rápido para suplir las necesidades del grupo de adultos mayores, que cada vez es mayor, trayendo consigo problemas sanitarios y de atención [20]. Por ello se hace evidente cada vez más grupos de apoyo social y de atención, donde los adultos mayores enfermos leves, enfermos graves, sanos y/o rutinarios intentan ayudarse entre ellos con soporte de pocos enfermeros, para no tener que llenar las camillas de un hospital o hacer largas filas para esperar unos resultados.

Los grupos de soporte social, han venido creciendo ya que el servicio de atención se ha vuelto más complejo de tomarlo en los centros hospitalarios o de atención, por lo cual este servicio se presta a domicilio, trayendo consigo múltiples beneficios como un servicio personalizado y un apoyo en casa, una realidad que beneficia a este grupo de adultos mayores. También estos grupos facilitan la capacitación de las personas encargadas del cuidado de los adultos mayores, que presentan alguna enfermedad crónica discapacitante o limitante, en los cuales se hace evidente la ayuda de otra persona que se debe capacitar u ofrecerles alternativas de conocimiento para que no presenten inconvenientes al momento del cuidado. Adicionalmente se debe tener un trato especial, donde la actividad del cuidado no afecte al cuidador socialmente, anímicamente o físicamente y así presentar una ayuda óptima con el paciente o familiar enfermo.

En Colombia, como país en vía de desarrollo, los cuidadores de enfermos crónicos en gran porcentaje son parte del núcleo familiar, lo cual le permite al enfermo crónico tener un apoyo constante en su cuidado, pero trayendo consigo consecuencias que afectan a dichos cuidadores, en diferentes aspectos de su calidad de vida haciendo plausible el desarrollo de programas de soporte social, interpretando estos como la interacción social, la guía para la retroalimentación, una ayuda tangible, un intercambio de información y capacitación, que facilite la comprensión de su rol como cuidador [21]

En todo el mundo los grupos de apoyo social se enfocan en su mayoría en una de las enfermedades características más relevantes y que determinan en gran porcentaje una población. Esto se debe a que los enfermos, son los que deben buscar ese apoyo, deben interesarse en conocer más a fondo cada detalle de su enfermedad. Estos grupos en todo el mundo colaboran entre sí para extender sus experiencias y conocimientos de manera que sea eficiente el trabajo y la cobertura sea mayor, esto se conocen como redes de apoyo, donde se busca una alianza para combatir las enfermedades. Estos grupos de soporte social a enfermedades se encargan de dar cursos, charlas, discusiones y demás recursos, que pueden proporcionar un conocimiento más a fondo de la enfermedad que padece la persona, una mayor ayuda a quienes lo cuidan y una satisfacción a quien la imparte.

Estos grupos tratan de una forma particular a cada persona, para lograr caracterizar de una manera más eficiente las enfermedades que padecen. Esto hace que los encargados del cuidado construyan una relación más íntima con la persona enferma, ayudando a buscar el camino más llevadero en su enfermedad. Por ello estos grupos se convierten en la parte formativa y social que la persona enferma necesita, no solamente porque se relaciona al profesional con el enfermo crónico, sino que también se crea ese intercambio de experiencias y esas relaciones entre enfermos crónicos.

Los países desarrollados en Europa y Norte América, han demostrado con experiencias como ha sido de efectivo este trabajo y como las personas con enfermedades ha superado sus limitaciones mentales y en gran parte han llevado su enfermedad de una manera más amable.

En Latinoamérica se pueden destacar varios países como focos de redes de apoyo social a las personas enfermas, tanto adultos mayores, como enfermos crónicos o pasantes, como son Argentina, Panamá, Guatemala y Colombia. Y en este último se pueden destacar las ciudades de Santa Martha, Cartagena, Barranquilla, Cúcuta, Ibagué, Girardot y Villavicencio [18].

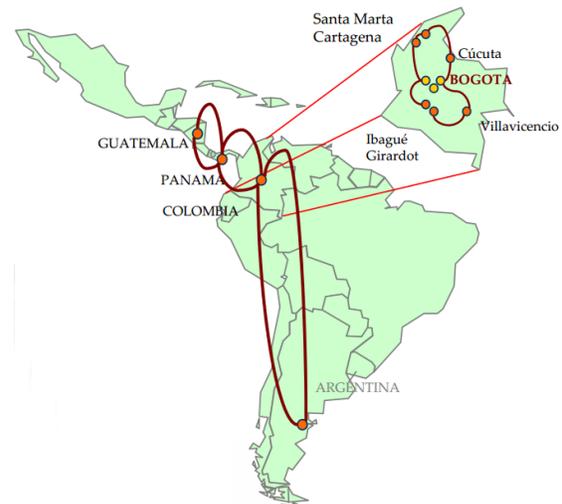

Fig 1. Mapa de redes sociales de Latinoamérica y Colombia [18].

### IV. SOPORTE SOCIAL BASADO TIC

La aplicación de las nuevas tecnologías para implantación de un mejor servicio y una mejor calidad de vida de las personas participes de los programas de soporte social ha venido presentando un creciente auge, lo cual también ha mejorado la cobertura y calidad de dichos programas; que de manera consecuente permite que poblaciones con poca accesibilidad a los programas de soporte social, puedan ser partícipes y beneficiarse de los mismos. Teniendo en cuenta el desarrollo tecnológico actual, es de resaltar en particular las TIC (tecnología de la información y las telecomunicaciones), que permite una interacción más directa con los usuarios y que ha tenido un gran avance en los últimos años.

Las TIC son herramientas con las que se escoge, organiza y promueve información, para fomentar el desarrollo de conocimientos y habilidades en las personas, y en este caso en particular a enfermos crónicos y sus cuidadores, de igual forma se ven aplicadas al soporte social permitiendo así ser un apoyo para los anteriormente nombrados. Teniendo en cuenta que las TIC permiten y facilitan los procesos de adaptación de cambios respecto comportamientos y estilos de vida que favorecen la salud física y mental tanto del enfermo crónico como su cuidador. [2], [22]-[24]. Así mismo se eliminan barreras de accesibilidad y altos costos del servicio. Estas TIC definen estrategias para conformar redes de apoyo y brindar un sistema de soporte social con mayor seguimiento, agilidad, versatilidad y cobertura, que a su vez integra conexiones a través de medios de comunicación y soporte en línea, facilitando así interacción permanente entre los pacientes con enfermedades crónicas, sus cuidadores familiares y el sistema de salud [2], [25].

Teniendo en cuenta lo anteriormente descrito se evidencia la necesidad de aplicación dichas tecnologías, que abarcan desde sistemas de teleformación (e-salud, e-learning, multimedia, …) hasta sistemas de comunicaciones (foros, videoconferencias, chat, …), que sean aplicadas al soporte social; considerando la importancia del cuidado enfocado hacia el enfermo crónico y el apoyo a su cuidador, demuestra la trascendencia y contribución del uso de TIC dentro del proceso de soporte social a este grupo en particular, proporcionando de manera directa ventajas para el progreso positivo, tanto del enfermo crónico y su convalecencia, como la tarea efectuada por su cuidador, mejorando su calidad de vida y contribuyendo a una mejor relación entre Cuidador y enfermo crónico. [2]

La aplicación de TIC como respuesta o apoyo para el desarrollo del soporte social, en diferentes grupos de personas en estado de cronicidad y sus cuidadores, ha tenido múltiples aplicaciones en distintos ámbitos, los cuales se han definido por medio de una revisión bibliográfica, gracias a las bases de datos proporcionadas por la Universidad de Cundinamarca de las cuales se hizo la selección de 65 artículos referentes al soporte social, de los cuales 31 se enfocan en el uso de las TIC como herramienta que sustenta el apoyo de estos programas. Con el fin de generar un panorama amplio de la inclusión de estas, en los programas de soporte social, dentro de la cuales cabe destacar distintos tipos del uso de estas, a modo de ejemplo el uso de la telefonía para la revisión periódica de los usuario de la redes [26]-[28], envío de información pertinente vía Email [26],[29] telemedición de signos vitales [30]-[32], videoconferencias [30], uso de blog's para difundir información [29], [33], [34], mensajería de texto con consejos prácticos [28], uso de páginas web [30], [33], [35], [36], encuestas virtuales

[29], [33], [36], intercambio de información a través bases de datos [31], [37], [38], elementos de medición implantables [32], diagnostico a distancia [38], entre otras formas de uso de las TIC, que se ven aplicadas al soporte social.

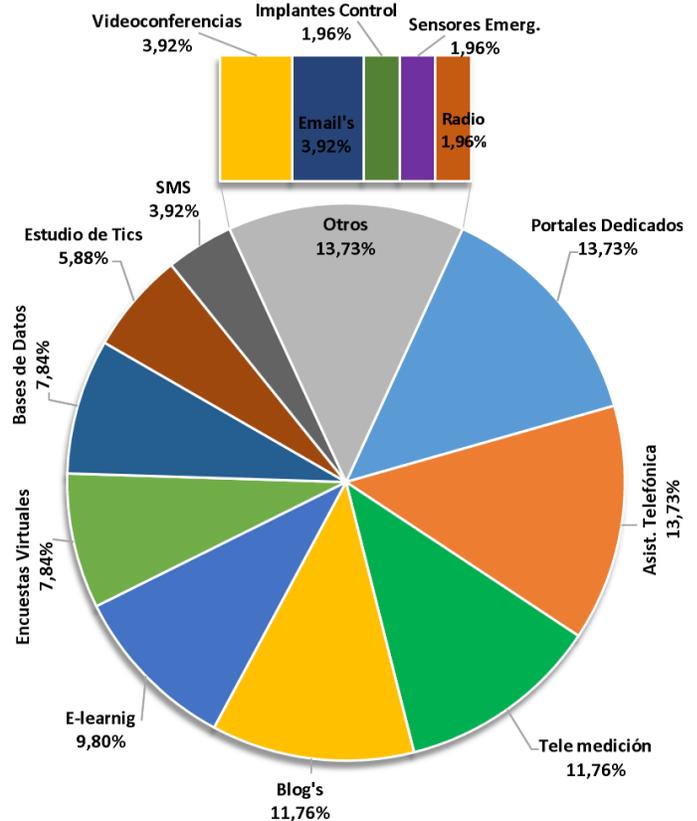

Fig 2. Niveles porcentuales de inclusión cada una de las herramientas TIC aplicadas al soporte social, definidas por la revisión bibliográfica.

La grafica anterior refleja el análisis de la base documental obtenida mediante el ejercicio bibliográfico (recolección de información, búsqueda en bases de datos, fichas bibliográficas generadas), en el cual se identificaron los tipos de tecnologías usadas dentro de las diferentes problemáticas, abarcadas en los papers, de la figura 2 podemos aprehender el porcentaje de inclusión de estas tecnologías (TIC) en el soporte social, mostrándonos una tendencia al uso de portales web dedicados, como una de las tecnologías más influyentes en el desarrollo del soporte social, no siendo menor el uso de la telefonía en estos programas, también cabe destacar que algunos soportes bibliográficos hacían en ocasiones alusión a varios tipos de herramientas tecnológicas, las cuales fueron implementadas para la solución de la problemática abordada por el mismo.

Para cada una de las tecnología adyacentes definidas por esta gráfica, se identifica una dispersión entre las magnitudes porcentuales calculadas, siendo esta exigua a comparación del dato inmediatamente menor, teniendo como valor recurrente un decaimiento del 1,96% en promedio, suscitando y confirmando así la poca diferencia anteriormente citada. De forma directa se pude evidenciar algunas de las herramientas tecnológicas que forman parte de este apoyo a los programas de soporte social.

Para efectos de este estudio, se tomó como parámetro cronológico una escala anual, comprendida en un intervalo de tiempo de 3 lustros, lo cual nos ayuda a determinar el desarrollo más reciente de estas tecnologías. Para ello se identificaron las fechas de publicación de cada documento, y de esta manera generar un análisis cualitativo, el cual se ve reflejado en la figura 3.

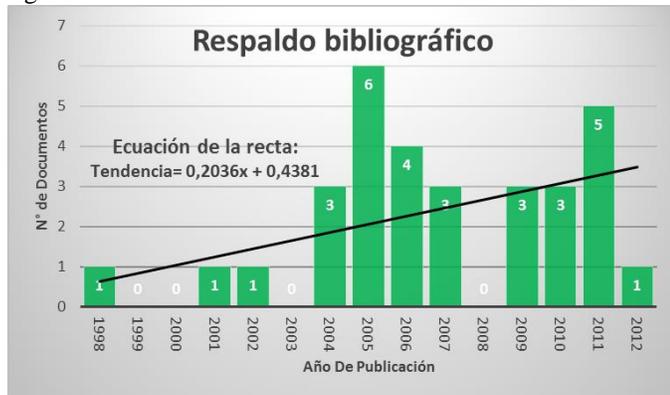

Fig 3. Número de Respaldos bibliográficos según el año de publicación.

La figura 3 nos presenta de manera directa, el año 2005 como el de mayor nivel de concentración, en cuanto al desarrollo bibliográfico se refiere, de manera agregada se identificó la línea de tendencia, usando el método de mínimos cuadrados, para identificar la recta más cercana a todos los puntos y así obtener una posible tendencia del uso de la TIC en el soporte social, la cual nos muestra que se encuentra en un aumento considerable. De forma paralela se identificó que en cuatro periodos, no hubo desarrollo de material investigativo acerca de esta temática, dentro de la muestra tomada como punto de partida, lo cual nos presenta una fluctuación en el constante desarrollo de dichas investigaciones.

Posterior a los análisis anteriores se realizó una distribución geográfica acerca de la concentración de cada uno de estos materiales investigativos, y de esta forma poder identificar el nivel de aprehensión del soporte social basado en TIC en el mundo.

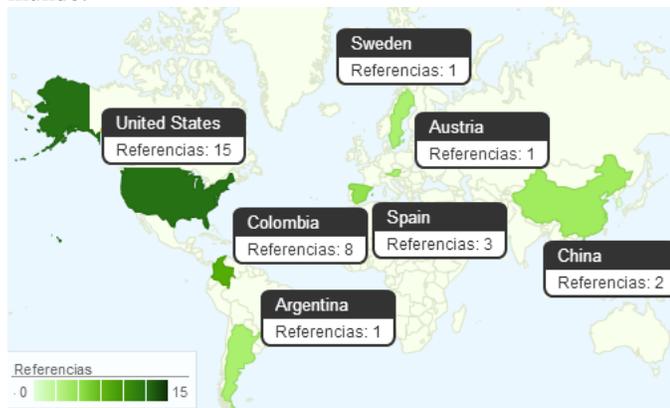

Fig 4. Mapa mundial de soporte social basado TIC, según el número de referencias estudiadas.

A nivel mundial se puede evidenciar que el país que se encuentra a la vanguardia en cuanto a la integración de TIC en el soporte social, es Estados Unidos puesto que capta 15 referencias de las 31 revisadas por el estudio, presentando así un 48,38% del total de estas, es de resaltar que en este caso el número de referencias usadas para definir el marco mundial, pude afectar la integración de otros países, así como puede favorecer la de otros, en este caso Colombia es uno de los cuales presenta una mayor resolución en el número de materiales investigativos. Lo cual nos permite hacer un análisis de estas herramientas y su uso a nivel nacional.

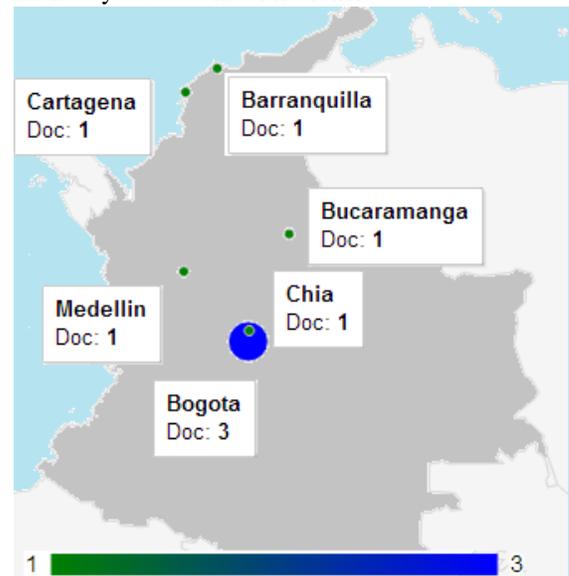

Fig 5. Mapa de Colombia acerca del soporte social basado TIC, según el número de referencias estudiadas.

La grafica determina como eje central con respecto a la distribución de los artículos a nivel nacional a Bogotá, como principal ciudad fomentadora de investigación y de material científico acerca de las herramientas TIC para el soporte social, seguida de no menor importancia la ciudad de Medellín, Chía, Bucaramanga, Barranquilla y Cartagena, las cuales auguran un desarrollo prospero en la temática y en los ejes del apoyo social, dependiendo los proyectos que se desarrollen y los recursos que se asignen para abarcar esta problemática social.

La distribución de los artículos de la revisión se enfoca en conocer el desarrollo de las herramientas utilizadas en la región y como se ha desenvuelto en el soporte social, para determinar cómo abarcar cada población y la óptima incursión de cualquier herramienta, así mismo determinar que herramienta optimiza todas las características de una población y de la cual se podrían aprovechar todos los recursos.

## V. DISCUSIÓN

La utilización de TIC en las actividades diarias, pueden traer consigo múltiples beneficios, ya que están diseñadas para ampliar el alcance de cualquier actividad en sus múltiples características, tales como comunicación, desplazamiento, tiempo, cobertura, etcétera y/o diferentes desventajas, dependiendo del objetivo a alcanzar con la aplicación de TIC. En el soporte social las TIC, pueden aportar en cuanto al apoyo social mayor flexibilidad, ya que por una parte no están regidas a la presencialidad y a todos los factores que esto atañe, como lo es el desplazamiento, la disponibilidad de tiempo, el acceso económico, acceso a la información, el oportuno diagnóstico o cuidado, todos estos pueden traer múltiples complicaciones que pueden bajar la calidad del servicio, provocar deserción en un número grande de integrantes de los grupos o en muchos casos la inconsistente asistencia que puede provocar vacíos informativos.

El aplicar TIC al soporte social, puede mejorar uno de los factores importantes cuando se quieren crear solidas redes de apoyo, el cual es la cobertura, debido a que por múltiples componentes y diferentes limitantes, la población del alcance del apoyo social no se concentra geográficamente en un punto central, por lo cual se puede excluir a personas interesadas en participar, que no pueden asistir a las actividades de estos nodos. Esta limitante puede verse caracterizada por varios factores, el cual en mayor porcentaje está sujeto a la economía, las poblaciones que no tienen acceso a un seguro prepagado o a una buen seguro médico, están de algún modo vulnerables, porque al llegar a un punto de longevidad, las enfermedades se hacen cada vez más recurrentes e inevitables y con la dificultad que tienen al acceso al diagnóstico y al cuidado, puede afectar la salud de una manera intransigente e irreversible. Aunque la atención fuera ligeramente accesible, la mayoría de la población vulnerable debe transportarse, para llegar al sitio donde se le presta atención a su salud, lo cual evidencia el papel que juega la economía y de la cual dependen otras actividades.

El no contar con el acceso a los recursos tecnológicos es otra consecuencia de la de la variedad de los niveles económicos de la población objeto, lo que implica que algunas personas en estado de cronicidad y sus cuidadores, no puedan tener acceso a los medios de comunicación o a las tecnologías que se estén usando en la aplicación de TIC, lo cual representa una gran barreara para el desarrollo de la actividad del soporte social. No solo la economía juega un papel importante, la disponibilidad de tiempo es un factor que debe tenerse en cuenta, debido a que cada persona desarrolla diferentes actividades que pueden cruzarse con la asistencia al servicio de apoyo. Por todos estos factores que limitan los grupos de apoyo las TIC, aparecen como una herramienta que proporcionaría el fácil desarrollo de grupos de apoyo sólidos y de amplios nodos que puedan abarcar regiones enteras.

Las TIC, se convierten en la salida a múltiples limitantes pero trae consigo también varias desventajas debido a que pude permitir una virtualización parcial de este apoyo, bloqueando una interacción personal con las respectivas relaciones, trayendo consigo el problema que aún se discute acerca de virtualizar servicios presenciales, que sin una respectiva guía o una utilización eficiente se puede convertir en obstáculo y/o hasta desmejorar el servicio prestado. Pero la virtualización también trae diferentes ventajas como es el hecho de la flexibilidad en cuanto a contenidos y horas de participación permitiendo así un acceso más completo y personalizado para cada usuario del programa.

Por todo esto las TIC, pueden aportar al servicio de atención en el cuidado optimo o simplemente desmejorarlo, por lo cual se deben tomar medidas como una buena caracterización de la población fundamental para acoplar los servicios de teleasistencia a los usuarios, de tal forma que se utilicen al máximo los recursos y que los pacientes enfermos crónicos y cuidadores de estos, se sientan a gusto y conformes que el nuevo servicio de asistencia.

La aplicación de dichas tecnologías de la información y las comunicaciones, deben ser tomadas como una herramienta orientada al apoyo del soporte social, mas no debe implicar la disminución del programa de soporte presencial, permitiendo así un trabajo en conjunto y generando de esta manera una solución más viable, efectiva y enfocada a esta población.